\documentclass[aps,pra,twocolumn,groupedaddress,amssymb,amsmath,showpacs]{revtex4}
\usepackage{dsfont}
\usepackage{bm}
\usepackage{bbm}
\usepackage[mathscr]{euscript}
\usepackage{amsmath}
\usepackage{amssymb}
\usepackage{amsthm}
\usepackage{amsfonts}
\usepackage{graphicx}
\usepackage{color}
\usepackage[colorlinks=true,urlcolor=blue,linkcolor=blue,citecolor=green]{hyperref}

\begin{document}

%\preprint{}

\title{Majorization Formulation of Uncertainty in Quantum Mechanics}


\author{M. Hossein Partovi}
\email[Electronic address:\,\,]{hpartovi@csus.edu}
%\thanks{}
\affiliation{Department of Physics and Astronomy, California State
University, Sacramento, California 95819-6041}


\date{\today}

\begin{abstract}
Heisenberg's uncertainty principle is formulated for a set of generalized measurements within the framework of majorization theory, resulting in a partial uncertainty order on probability vectors that is stronger than those based on quasi-entropic measures.  The theorem that results from this formulation guarantees that the uncertainty of the results of a set of generalized measurements without a common eigenstate has an inviolable lower bound which depends on the measurement set but not the state.  A corollary to this theorem yields a parallel formulation of the uncertainty principle for generalized measurements based on quasi-entropic measures.  Optimal majorization bounds for two and three mutually unbiased bases in two dimensions are calculated.  Similarly, the leading term of the majorization bound for position and momentum measurements is calculated which provides a strong statement of Heisenberg's uncertainty principle in direct operational terms.  Another theorem provides a majorization condition for the least uncertain generalized measurement of a given state with interesting physical implications.

\end{abstract}


%\keywords{}

\maketitle

%\section{}
%\subsection{}
%\subsubsection{}

% If in two-column mode, this environment will change to single-column
% format so that long equations can be displayed. Use
%\begin{widetext}
%\end{widetext}

\section{Introduction}
Heisenberg's uncertainty principle codifies certain inherent limitations on the simultaneous knowledge of observables in the microscopic realm, and as such constitutes one of the conceptual pillars of quantum theory \cite{HEI}. The limitations implied by the uncertainty principle played an important role in the celebrated Bohr-Einstein debates and the formative years of quantum theory, and have  since served as a source of insight on the structure and behavior of microscopic systems.  Having recognized that the non-commutativity of a pair of observables implies an irreducible indeterminacy in the simultaneous knowledge of their values, Heisenberg presented semi-quantitative arguments to establish a universal lower bound of the order of Planck's constant for the product of their uncertainties.  Heisenberg's arguments were subsequently formulated in a mathematically precise manner by Kennard \cite{KEN} and extended by Robertson \cite{ROB} and Schr\"{o}dinger \cite{SCH}, all of whom adopted the square root of variance as the measure of uncertainty.  The variance formulation of the uncertainty principle, which is the one
familiar from textbook accounts, is often useful when applied to canonically conjugate observables for estimating spectral
and structural properties of microsystems.

More than half a century later it was realized that entropy is a more effective measure for capturing the information
theoretical aspects of uncertainty, especially when applied to noncanonical observables.  The key idea was to quantify uncertainty as the information associated
with the probabilities of measurement outcomes rather than the variance in the values of the measured observable. In a seminal paper \cite{DEU},
Deutsch argued that such a measure would be superior to the traditional one when dealing with observables
of finite rank, and developed an entropic formulation of the uncertainty principle based on an inviolable lower bound for
the sum of Shannon entropies associated with projective measurements of noncommuting observables.  He also emphasized that, in contrast to the variance formulation in its generalized form, the irreducible lower bound in the entropic formulation depends on the observables but not the state.

In a subsequent work, this author showed that an extension of Deutsch's formulation to infinite-rank observables such as position and momentum, whether
discrete or continuous, entails a nontrivial consideration of the resolution of the measuring device \cite{HP1}. The resulting
entropic measure was a more realistic description of uncertainty for canonical observables than the variance formulation, and in contrast to the latter, was mathematically well defined for any state of the system and an arbitrary set of observables, even if infinite-rank. Consequently, while Deutsch's primary objective was to deal with the shortcomings of the variance formulation when applied to discrete spectra and finite-dimensional observables, the extension to continuous spectra such as position and momentum turned out to be quite potent as well.

The entropic formulation has since been refined by deriving sharper bounds \cite{BI1, KMU} and extended by introducing mutually unbiased observables \cite{KMU}, generalized measurements \cite{SHA},
alternative entropy functions such as those of Tsallis \cite{RAJ} and R\'{e}nyi \cite{BIA}, and other innovations \cite{SUR}.
Its applications include the quantum formulation of Jaynes' maximum entropy method and time-energy uncertainty relations
\cite{HP2}, and more recently quantum cryptography, information locking, and entanglement detection \cite{SUR, GAM}.  In particular, the recent applications to quantum information and entanglement theory clearly underscore the importance of basing uncertainty considerations on probability vectors resulting from measurements rather than values of measured observables.  It should also be noted here that the idea of entropy as a measure of uncertainty has an interesting earlier history \cite{PRE}.

In this paper we develop a majorization formulation of the uncertainty principle.  Majorization provides a partial order on
probability vectors which characterizes the degree of their disorder, or uncertainty, and is naturally suited for application
to measurement results.  It is based on the intuitive but surprisingly powerful notion that a probability vector which is a
mixture of the permutations of another is more disordered.  This simple condition gives rise to a measure of uncertainty that is
more discriminating than any measure based on a symmetric, concave function defined on probability vectors.  We shall refer to
this class of measures, which includes the Shannon and Tsallis (but not R\'{e}nyi) entropies, as ``quasi-entropic.'' Thus the
majorization order implies any quasi-entropic order, but not vise versa \cite{MAO}.  Indeed the main theorem of the present formulation directly implies the existence of a class of scalar formulations of the uncertainty principle for generalized measurements based on such quasi-entropic functions.  This generalization serves, inter alia, to extend the standard (Shannon) entropic measure of uncertainty to a set of generalized measurements.

It is important to note here that quasi-entropic order is total and based on a single inequality, whereas the majorization order is partial (i.e., not every pair of probability vectors can be ordered) and rests on $N-1$ inequalities, where $N$ is the number of nonzero entries in the less uncertain probability vector under comparison (see \S IIB for details). Consequently, while the majorization order is equivalent to quasi-entropic ones for $N \leq 2$, it becomes progressively more stringent with increasing $N$.  It is this feature of majorization, namely matching the complexity of the probe to that of the object, that makes it particularly effective as a comparator of disorder.  It should therefore come as no surprise that some of the most fundamental results of quantum information and entanglement theory are based on majorization relations \cite{MAJ}.  We believe that its use in the present formulation likewise serves to extend the reach and power of the uncertainty principle.

The rest of this paper is organized as follows: In \S IIA we describe measurement types and establish notation, in \S IIB we introduce the relevant elements of majorization theory, and in \S IIC we relate concepts of measurement uncertainty to majorization relations.  In \S IIIA and IIIB we establish the central result of this paper, Theorem 1, parts A and B, and state the physical content of the uncertainty principle in majorization terms.  In \S IIIC, we construct the class of quasi-entropic measures of uncertainty as a corollary to Theorem 1.  We apply Theorem 1 to mutually unbiased observables in \S IVA and IVB, and to position and momentum in \S V.  In \S VI we establish the least uncertain measurement of a quantum state in Theorem 2, and state its relation to the von Neumann entropy and spectrum of the state.  We conclude the paper with a few remarks in \S VII.
\section{measurement uncertainty and majorization}
We start by introducing the concepts and methods that underlie the majorization formulation of uncertainty in the following subsections.
\subsection{Measurements}
To establish our notation and nomenclature, we start by defining types of measurement.  A generalized measurement is defined by
a set of measurement operators $\{ {\hat{\mathfrak{M}}}_{\alpha} \}$ subject to the completeness condition ${\sum}_{\alpha}
{\hat{\mathrm{E}}}_{\alpha} = \hat{\mathbbm{1}}$, where ${\hat{\mathrm{E}}}_{\alpha}={\hat{\mathfrak{M}}}^{\dagger}_{\alpha} {\hat{\mathfrak{M}}}_{\alpha} $
is called a measurement element and the index $\alpha$ identifies the possible measurement outcomes \cite{NCH}.   Note that each measurement element is a bounded, positive, self-adjoint operator whose norm cannot exceed unity.  The
probability that outcome $\alpha$ turns up in a measurement of the state $\hat{\rho}$ is given by the Born rule
$\mathscr{P}_{\alpha}(\hat{\rho})=\textrm{tr} [\hat{\mathrm{E}}_{\alpha} \hat{\rho} ]$, with the post-measurement state given by
${\hat{\rho}'}_{\alpha}={\hat{\mathfrak{M}}}_{\alpha} \hat{\rho} \, {\hat{\mathfrak{M}}}^{\dagger}_{\alpha}/\mathscr{P}_{\alpha}$ \cite{state}. If $\hat{\rho}$ is not a
pure state, then ${\hat{\rho}'}_{\alpha}$ will not in general be pure unless ${\hat{\mathfrak{M}}}_{\alpha}$ is of rank $1$ \cite{rank}. A measurement is
rank-1 if every measurement operator is of rank $1$.  Rank-1 measurements are thus seen to have the highest resolution among generalized measurements in the sense that the range of their measurement operators consists of a single pure state.

A generalized measurement can always be considered as a restriction of a more basic type of measurement, namely a \textit{projective} measurement performed on an enlarged system, to the system under generalized measurement \cite{NCH}.  A projective measurement is commonly associated with an observable of the system, which would be represented by some self-adjoint operator $\hat{M}$.  Such a measurement entails a partitioning of the spectrum of $\hat{M}$ into a
collection of subsets $\{ {b}_{\alpha}^{M} \}$ called \textit{bins} \cite{HP1}.  This partition induces a corresponding
partition of the Hilbert space into orthogonal subspaces with the corresponding projection operators being the measurement
operators \cite{proj}.  Thus for projective measurements, ${\hat{\mathfrak{M}}}_{\alpha}^{M}={\hat{\Pi}}_{\alpha}^{M}$, where $ {\hat{\Pi}}_{\alpha}^{M} $ is the projection operator onto
the subspace corresponding to ${b}_{\alpha}^{M}$.  We call a projective measurement \textit{maximal} if each bin consists of a
single point of the spectrum of the measured observable.  Note that a maximal measurement of an observable with a non-degenerate
spectrum, the type usually described in textbook accounts, is rank-1. It is important to understand that physically realizable
measurements are limited to a finite set of outcomes so that any measurement of an observable with a continuous spectrum such as
position or momentum, or even an infinite discrete spectrum such as the energy of a harmonic oscillator, must necessarily involve infinite-rank measurement operators and cannot be maximal \cite{HP1}.
Equivalently, only systems describable by finite-dimensional Hilbert spaces (such as spin systems) admit maximal measurements.
\subsection{Majorization relations}
We now turn to a brief introduction to the basics of majorization relations, a topic that has found important applications in quantum information and entanglement theory \cite{MAO,MAJ}.  With every vector $\lambda$ we associate another vector ${\lambda}^{\downarrow}$ which is obtained from $\lambda$ by arranging the components of the latter in a descending (i.e., nonincreasing) order.  Then, given a pair of vectors ${\lambda}^{1}$ and ${\lambda}^{2}$, ${\lambda}^{1}$ is said to be majorized by ${\lambda}^{2}$ and written ${\lambda}^{1}\prec{\lambda}^{2}$ if
${\sum}_{i}^{j} {\lambda}^{1\downarrow}_{i} \leq {\sum}_{i}^{j} {\lambda}^{2\downarrow}_{i} $ for $j=1,2, \ldots d$, where $d$ is the larger of the
two dimensions and trailing zeros are added where needed.  An equivalent but intuitively more compelling
definition is that ${\lambda}^{1}\prec{\lambda}^{2}$ if and only if ${\lambda}^{1}$ equals a mixture of permutations of
${\lambda}^{2}$.  As stated earlier, the majorization relation is a \textit{partial} order, i.e., that not every two vectors are
comparable under majorization. Note that this is not a shortcoming of majorization, but rather a
consequence of its more rigorous protocol for ordering uncertainty.  Indeed as mentioned in \S I, for any function $F(\lambda)$ of the quasi-entropic type like Shannon and Tsallis entropies, ${\lambda}^{1} \prec {\lambda}^{2}$ implies $F({\lambda}^{1}) \geq
F({\lambda}^{2})$, but not conversely \cite{MAO}.  On the other hand, if for \textit{every} such function $F(\lambda)$ we have
$F({\lambda}^{1}) \geq F({\lambda}^{2})$, then ${\lambda}^{1} \prec {\lambda}^{2}$.  Clearly, the majorization relation as a
comparator of disorder is stronger than any single quasi-entropic measure, and in a sense is equivalent to all such measures taken collectively.

Another concept needed in the following is that of the \textit{infimum} of a set of $N$ vectors, defined as the vector that is
majorized by every element of the set and in turn majorizes any vector with that property \cite{HP3}.   The \textit{supremum} is
similarly defined as the vector that majorizes every element of the set and is in turn majorized by any vector with that
property \cite{note1}.  To construct these extremal vectors, we consider a vector ${\mu}^{inf}$ with components
\begin{align}
{\mu}_{0}^{inf}=0&,\,\,\,{\mu}_{j}^{inf}=\min \big ( {\sum}_{i=1}^{j} {\lambda}^{1\downarrow}_{i},{\sum}_{i=1}^{j} {\lambda}^{2\downarrow}_{i},\ldots, \nonumber \\
&{\sum}_{i=1}^{j}{\lambda}^{N\downarrow}_{i} \big ), \,\,\, 1 \leq j \leq {d}_{max}, \label{1}
\end{align}
where ${d}_{max}$ is the largest dimension found in the set \cite{note11}.  The desired infimum is then given by
\begin{equation}
{\lambda}^{inf}_{i}={[\inf({\lambda}^{1},{\lambda}^{2}, \ldots, {\lambda}^{N})]}_{i}={\mu}_{i}^{inf}-{\mu}_{i-1}^{inf}, \label{1.5}
\end{equation}
where $1 \leq i \leq {d}_{max}$.  One can show that ${\lambda}^{inf}$ as given by Eq.~(\ref{1.5}) is a descending sequence \cite{des}.  While the construction given in Eq.~(\ref{1.5}) guarantees that ${\lambda}^{inf}$ is majorized by every element of the set, the descending property guarantees that any other vector with that property is in turn majorized by  ${\lambda}^{inf}$.

A parallel construction to the above with ``min'' in Eq.~(\ref{1}) replaced with ``max'' yields a sequence that majorizes every element of the set, but one that does not necessarily emerge in a descending order and may therefor fail to be majorized by any other sequence that has the same property, as required.  In such a case, the sequence so obtained, $\tilde{\lambda}$, can be modified by a ``flattening'' process that, while maintaining the property of majorizing every element of the set, culminates in a sequence that is descending as well.  The flattening process starts with $\tilde{\lambda}$, and for every pair of components violating the descending property, i.e., ${\tilde{\lambda}}_{i+1} > {\tilde{\lambda}}_{i}$, replaces the pair by their mean such that the updated elements are ${\tilde{\lambda }'}_{i}={\tilde{\lambda}'}_{i+1}=({\tilde{\lambda}}_{i}+{\tilde{\lambda}}_{i+1})/2$.  This process of ``flattening'' is then continued until a descending sequence corresponding to the supremum ${\lambda}^{sup}$ is obtained \cite{code}.
\subsection{Uncertainty}
We are now in a position to characterize uncertainty by means of majorization relations.  The probability vector
$\mathscr{P}^{X}(\hat{\rho})$ resulting from a measurement $\mathrm{X}$ on a state $\hat{\rho}$ is said to be \textit{uncertain} if it is
majorized by $\mathcal{I}=(1,0,\ldots,0)$ but not equal to it.  As such, $\mathscr{P}^{X}(\hat{\rho})$ is said to be strictly
majorized by $\mathcal{I}$ and written $\mathscr{P}^{X}(\hat{\rho})\prec \prec \mathcal{I}$. Similarly, $\hat{\rho}$ is more uncertain with
respect to measurement $\mathrm{X}$ than with respect to $\mathrm{Y}$ if $\mathscr{P}^{\mathrm{X}}(\hat{\rho}) \prec
\mathscr{P}^{\mathrm{Y}}(\hat{\rho})$. Furthermore, we define the joint uncertainty of a pair of measurements $\mathrm{X}$ and
$\mathrm{Y}$ by means of the outer product $\mathscr{P}^{\mathrm{X}}\otimes\mathscr{P}^{\mathrm{Y}}$, i.e., $\mathscr{P}_{\alpha
\beta}^{\mathrm{X} \oplus \mathrm{Y}}=\mathscr{P}^{\mathrm{X}}_{\alpha} \mathscr{P}^{\mathrm{Y}}_{\beta}$.  Since
$H(\mathscr{P}^{\mathrm{X}}\otimes \mathscr{P}^{\mathrm{Y}})=H(\mathscr{P}^{\mathrm{X}})+H(\mathscr{P}^{\mathrm{Y}})$, where $H(\cdot)$ is the Shannon entropy function, this definition is seen to be consistent with its entropic counterpart.   As stated earlier, $\mathscr{P}^{\mathrm{X}} \prec \mathscr{P}^{\mathrm{Y}}$ implies $H(\mathscr{P}^{\mathrm{X}}) \geq
H(\mathscr{P}^{\mathrm{Y}})$ but not conversely.  Note that the foregoing definitions naturally extend
to an arbitrary number of states and measurements.

It is worth repeating here that the partial nature of the majorization order implies that not all measurement results are comparable under our uncertainty order.  While a total uncertainty order such as given by the entropic formulation would be simpler to deal with, it would also embody less information as an averaged out feature of a probability vector, especially for vectors of high disorder.  A concomitant of partial order is that the infimum or supremum of a set of probability vectors will in general not be a member of the set.  In other words, a set of probability vectors does not in general have a least uncertain element, although it may have multiple elements that are not more uncertain than any other probability vector \cite{maximal}.  While one may wish to do away with these unfamiliar features, they are nevertheless a small price to pay for a high-resolution comparator of uncertainty.
\section{majorization formulation of the uncertainty principle}
The information theoretical expression of the uncertainty principle may be stated as the requirement that the information available from a set of measurements performed on a system is subject to an irreducible level of uncertainty unless the measurements have a common eigenstate.  In the latter case, the outcome of every measurement would be certain and unique, and the information so obtained would be complete for the given set of measurements.  An important additional requirement is that the said irreducible level of uncertainty be a property of the measurement set and not depend on the state of the system.  The entropic formulation implements these requirements by identifying the sum of the entropies associated with the measurements in the set as the measure of their joint uncertainty \cite{DEU}.  The majorization formulation, by contrast, relies on the outer product of probability vectors resulting from the measurements as representative of their joint uncertainty.  An example of this definition for a set of two measurements was stated in \S IIC.  The corresponding irreducible level of uncertainty is then defined to be the supremum of such outer products as all possible states are considered.  This supremum is then guaranteed to be less uncertain than the probability vector resulting from the measurement of any possible state of the system.  As also noted in \S IIC, the Shannon entropy of the outer product of a set of probability vectors equals the sum of their entropies, which guarantees that the two formulations are consistent.  It also implies that a formulation of the uncertainty principle in terms of majorization implies a parallel entropic formulation, as will be demonstrated in \S IIIC.

In the following we will establish a theorem that embodies the above statements in a mathematically precise manner.  We will treat the cases where measurement elements have discrete spectra and behave similarly to Hermitian matrices acting on finite-dimensional spaces, namely \textit{compact} measurement elements, separately from other cases such as canonical observables and continuous spectra \cite{ReSi}.  The reason for this separate treatment is simply that the very notion of an eigenstate may lose proper mathematical meaning for non-compact measurement elements \cite{eig}.  Needless to say, the physical content of the uncertainty principle is unaffected by these mathematical details.
\subsection{Compact measurement elements}
Here we consider a measurement set whose elements are compact.  Since measurement elements are bounded, positive, self-adjoint operators, the condition of compactness will guarantee that they only have a discreet spectrum consisting of positive eigenvalues, each with a finite multiplicity, except possibly for a clustering of eigenvalues near zero \cite{ReSi}.  Since we will primarily be concerned with eigenvalues away from zero, compact elements essentially behave as positive Hermitian matrices (of finite dimension) for our purposes.  Note, however, that while finite-rank measurement elements are necessarily compact, the converse is not true.  Note also that most discussions of uncertainty in the literature deal with finite-rank, hence compact, elements, and often just rank-1 projection operators as in the case of maximal projective measurements.

\textbf{Theorem 1A.}  Let $\{ \mathscr{P}^{\mathrm{X}}(\hat{\rho}), \mathscr{P}^{\mathrm{Y}}(\hat{\rho}), \ldots, \mathscr{P}^{\mathrm{Z}}(\hat{\rho}) \}$ be the set of probability vectors resulting from a set of generalized measurements $\{ \mathrm{X}, \mathrm{Y}, \ldots, \mathrm{Z} \}$  with compact elements on the state $\hat{\rho}$. Then
\begin{equation}
\mathscr{P}^{\mathrm{X}}(\hat{\rho})\otimes \mathscr{P}^{\mathrm{Y}}(\hat{\rho}) \otimes \ldots \otimes \mathscr{P}^{\mathrm{Z}}(\hat{\rho}) \prec
{\mathscr{P}}_{sup}^{\mathrm{X} \oplus \mathrm{Y} \oplus \ldots \oplus \mathrm{Z}} \prec \prec \mathcal{I}, \label{2}
\end{equation}
where
\begin{equation}
{\mathscr{P}}_{sup}^{\mathrm{X} \oplus \mathrm{Y} \oplus \ldots \oplus \mathrm{Z}}=
{\sup}_{\hat{\rho}} [\mathscr{P}^{\mathrm{X}}(\hat{\rho})\otimes \mathscr{P}^{\mathrm{Y}}(\hat{\rho}) \otimes \ldots \otimes
\mathscr{P}^{\mathrm{Z}}(\hat{\rho})] , \label{2.1}
\end{equation}
unless the measurement elements $\{ \hat{\mathrm{E}}^{\mathrm{X}}, \hat{\mathrm{E}}^{\mathrm{Y}}, \ldots, \hat{\mathrm{E}}^{\mathrm{Z}} \}$ have a common eigenstate in which case ${\mathscr{P}}_{sup}^{\mathrm{X} \oplus \mathrm{Y} \oplus \ldots \oplus \mathrm{Z}} = \mathcal{I}$.

It is important to note that ${\mathscr{P}}_{sup}^{\mathrm{X} \oplus \mathrm{Y} \oplus \ldots \oplus \mathrm{Z}}$ depends on the measurement set but is independent of the state $\hat{\rho}$.  Note also that as the supremum of all possible measurement outcomes for the measurement set, it is the probability vector that sets the irreducible lower bound to uncertainty for the set.  As such, it is analogous to variance or entropic lower bounds for existing formulations of the uncertainty principle.  Unlike the scalar bounds of the variance and entropic formulations, however, ${\mathscr{P}}_{sup}^{\mathrm{X} \oplus \mathrm{Y} \oplus \ldots \oplus \mathrm{Z}}$ is in general a vector quantity whose dimension is variable and grows with the complexity of the measurement set.  In addition, ${\mathscr{P}}_{sup}^{\mathrm{X} \oplus \mathrm{Y} \oplus \ldots \oplus \mathrm{Z}}$ is in general not realizable on any state of the system \cite{note1} except under special conditions.  In other words, there is generally no such thing as a ``minimum uncertainty state'' within the majorization framework, as alluded to in \S IIC.  An important exception to this statement is the special case of zero uncertainty for which ${\mathscr{P}}_{sup}^{\mathrm{X} \oplus \mathrm{Y} \oplus \ldots \oplus \mathrm{Z}}= \mathcal{I}$, signaling the existence of a common eigenstate for the measurement elements as asserted by Theorem 1A.  We will illustrate these and other properties for three archetypal cases in \S IV and V.

To establish Theorem 1A, we need to show that ${\mathscr{P}}_{sup}^{\mathrm{X} \oplus \mathrm{Y} \oplus \ldots \oplus
\mathrm{Z}}$ is strictly majorized by $\mathcal{I}$, equivalently that its largest component is strictly less than unity, if the
measurement elements do not have a common eigenstate.  Suppose, on the contrary, that the largest component of ${\mathscr{P}}_{sup}^{\mathrm{X} \oplus \mathrm{Y} \oplus \ldots \oplus \mathrm{Z}}$ does equal unity while the measurement elements in the set do not possess a common eigenstate.  Then there must exist a set of indices $({\alpha}^{*}, {\beta}^{*}, \ldots, {\gamma}^{*})$ such that ${\sup}_{{\rho}} [\mathscr{P}^{\mathrm{X}}_{{\alpha}^{*}}(\hat{\rho})\mathscr{P}^{\mathrm{Y}}_{{\beta}^{*}}(\hat{\rho}) \ldots
\mathscr{P}^{\mathrm{Z}}_{{\gamma}^{*}}(\hat{\rho})]=1$, or in terms of measurement elements,
\begin{equation}
{\sup}_{{\rho}} [\textrm{tr}({\hat{\mathrm{E}}}^{X}_{{\alpha}^{*}} \hat{\rho}) \textrm{tr}({\hat{\mathrm{E}}}^{Y}_{{\beta}^{*}} \hat{\rho}) \ldots \textrm{tr}({\hat{\mathrm{E}}}^{Z}_{{\gamma}^{*}} \hat{\rho})]=1.  \label{2.2}
\end{equation}

Since the elements of each measurement are positive Hermitian operators whose sum equals the identity operator, e.g. ${\sum}_{\alpha} {\hat{\mathrm{E}}}^{X}_{\alpha} =\hat{ \mathbbm{1}}$, they must all be bounded operators with norms not exceeding unity, i.e. $\parallel {\hat{\mathrm{E}}}^{X}_{\alpha} \parallel \leq 1$ for every $\alpha$, with similar conditions for all other measurement elements in the set.  But this implies that $\textrm{tr} (\hat{\mathrm{E}}_{{\alpha}^{*}}^{X}{\hat{\rho}}) \leq 1$, with similar conditions for all measurement elements in the set.   Consequently, the only way Eq.~(\ref{2.2}) can be satisfied under the stated constraint on the norms is that (a) each measurement element is of unit norm, and (b) each trace term in Eq.~(\ref{2.2}) equals unity, i.e.,
\begin{equation}
{\sup}_{{\rho}}\,\textrm{tr} (\hat{\mathrm{E}}_{{\alpha}^{*}}^{X}{\hat{\rho}}, \hat{\mathrm{E}}_{{\beta}^{*}}^{Y}{\hat{\rho}}, \ldots,  \hat{\mathrm{E}}_{{\gamma}^{*}}^{Z}{\hat{\rho}})=(1, 1,\ldots,1). \label{2.3}
\end{equation}
Note that by probability conservation, if a subscript on an entry in the left-hand side of Eq.~(\ref{2.3}) is changed, the corresponding entry on the right-hand side must vanish.  Physically, Eq.~(\ref{2.3}) implies the existence of states for which the outcome of every measurement is essentially determinate, and the joint results are basically without any uncertainty, since the outcome of the measurement set $ \{ \mathrm{X}, \mathrm{Y}, \ldots, \mathrm{Z} \}$  will very nearly all be events in those measurement ``bins'' that correspond to the subscript set $({\alpha}^{*}, {\beta}^{*}, \ldots, {\gamma}^{*})$.

Mathematically, on the other hand, we note that the measurement elements in Eq.~(\ref{2.3}) are positive, compact Hermitian operators of unit norm, with discrete eigenvalues and corresponding eigenfunctions that are complete.  In addition, all nonzero eigenvalues have finite multiplicity.  We may therefore conclude that there exists a state ${\hat{\rho}}^{*}$ that realizes the equalities of Eq.~(\ref{2.3}), i.e., that $\textrm{tr}(\hat{\mathrm{E}}_{{\alpha}^{*}}^{X}{\hat{\rho}}^{*})= \textrm{tr}(\hat{\mathrm{E}}_{{\beta}^{*}}^{Y}{\hat{\rho}}^{*})= \ldots=  \textrm{tr}(\hat{\mathrm{E}}_{{\gamma}^{*}}^{Z}{\hat{\rho}}^{*})=1$, and that ${\hat{\rho}}^{*}$ is a common eigenstate of the measurement elements $\{ \hat{\mathrm{E}}_{{\alpha}^{*}}^{X}, \hat{\mathrm{E}}_{{\beta}^{*}}^{Y}, \ldots, \hat{\mathrm{E}}_{{\gamma}^{*}}^{Z} \}$ with eigenvalues unity (and of all other elements with eigenvalue zero) \cite{note12}.  However, this conclusion contradicts our starting assumption, thereby completing the proof of Theorem 1A.
\subsection{Non-compact measurement elements}
In the foregoing paragraph we used the compactness property of the measurement elements to deduce the existence of a common eigenstate for them from Eq.~(\ref{2.3}).  In the general case where non-compact elements may be present, the measurement elements may not even have properly defined eigenstates or eigenvalues, common or otherwise \cite{eig}.  Of course the physical content of the uncertainty principle is still captured by Eq.~(\ref{2.3}).  We will therefore use the physically equivalent notion of an \textit{approximate} eigenstate in this case, defined as follows: If for some number $a$ and any $\epsilon >0$ there exists a state $\psi(\epsilon)$ such that $\parallel (\hat{A}-a )\psi(\epsilon) \parallel < \epsilon$, then $\psi(\epsilon)$ is said to be an approximate eigenstate of $\hat{A}$.

Clearly, Eq.~(\ref{2.3}) implies the existence of an approximate common eigenstate for the measurement elements therein, thus yielding the desired result.  Nevertheless, it is useful to develop a separate formulation and an alternative proof for this general case, especially with a view to deriving majorization bounds for position and momentum measurements in \S V.

\textbf{Theorem 1B.}  Let $\{ \mathscr{P}^{\mathrm{X}}(\hat{\rho}), \mathscr{P}^{\mathrm{Y}}(\hat{\rho}), \ldots, \mathscr{P}^{\mathrm{Z}}(\hat{\rho}) \}$ be the set of probability vectors resulting from a set of generalized measurements $\{ \mathrm{X}, \mathrm{Y}, \ldots, \mathrm{Z} \}$ on the state $\hat{\rho}$.  Then Eq.~(\ref{2}) of Theorem 1A holds unless the measurement elements $\{ \hat{\mathrm{E}}^{\mathrm{X}}, \hat{\mathrm{E}}^{\mathrm{Y}}, \ldots, \hat{\mathrm{E}}^{\mathrm{Z}} \}$ have an approximate common eigenstate in which case ${\mathscr{P}}_{sup}^{\mathrm{X} \oplus \mathrm{Y} \oplus \ldots \oplus \mathrm{Z}} = \mathcal{I}$.

To establish this result, we first note that, by definition,
\[\parallel { \hat{\mathrm{E}}^{\mathrm{X}}_{\alpha}+ \hat{\mathrm{E}}^{\mathrm{Y}}_{\beta}+ \ldots + \hat{\mathrm{E}}^{\mathrm{Z}}_{\gamma}} \parallel \geq
\textrm{tr}[({ \hat{\mathrm{E}}^{\mathrm{X}}_{\alpha}+ \hat{\mathrm{E}}^{\mathrm{Y}}_{\beta}+ \ldots + \hat{\mathrm{E}}^{\mathrm{Z}}_{\gamma}}) \hat{\rho}] \]
for any (normalized) density operator $\hat{\rho}$.  The right-hand side of this inequality is, by definition, equal to $[\mathscr{P}^{\mathrm{X}}_{{\alpha}}(\hat{\rho})+\mathscr{P}^{\mathrm{Y}}_{{\beta}}(\hat{\rho})+ \ldots +
\mathscr{P}^{\mathrm{Z}}_{{\gamma}}]$, which is a sum of non-negative numbers.  Since the arithmetic mean of a set of non-negative numbers is never exceeded by their geometric mean, we have the inequality
\[ [\mathscr{P}^{\mathrm{X}}_{{\alpha}}(\hat{\rho})+\mathscr{P}^{\mathrm{Y}}_{{\beta}}(\hat{\rho})+ \ldots +
\mathscr{P}^{\mathrm{Z}}_{{\gamma}}]/n \geq  {[\mathscr{P}^{\mathrm{X}}_{{\alpha}}(\hat{\rho})\mathscr{P}^{\mathrm{Y}}_{{\beta}}(\hat{\rho}) \ldots
\mathscr{P}^{\mathrm{Z}}_{{\gamma}}]}^{1/n} \]

Combining the above pair of inequalities, we arrive at the important conclusion that
\begin{equation}
\mathscr{P}^{\mathrm{X}}_{{\alpha}}(\hat{\rho})\mathscr{P}^{\mathrm{Y}}_{{\beta}}(\hat{\rho}) \ldots \mathscr{P}^{\mathrm{Z}}_{{\gamma}}(\hat{\rho}) \leq
{(\parallel { \hat{\mathrm{E}}^{\mathrm{X}}_{\alpha}+ \hat{\mathrm{E}}^{\mathrm{Y}}_{\beta}+ \ldots + \hat{\mathrm{E}}^{\mathrm{Z}}_{\gamma}} \parallel/n)}^{n}, \label{2.4}
\end{equation}
where $n$ is the number of measurements in the set.

At this point we follow the proof of Theorem 1A by assuming, contrary to Theorem 1B, that ${\mathscr{P}}_{sup}^{\mathrm{X} \oplus \mathrm{Y} \oplus \ldots \oplus \mathrm{Z}}= \mathcal{I}$ while the measurement elements do not possess an approximate common eigenstate.  But then there must exist a set of indices $({\alpha}^{*}, {\beta}^{*}, \ldots, {\gamma}^{*})$ such that ${\sup}_{{\rho}} [\mathscr{P}^{\mathrm{X}}_{{\alpha}^{*}}(\hat{\rho})\mathscr{P}^{\mathrm{Y}}_{{\beta}^{*}}(\hat{\rho}) \ldots \mathscr{P}^{\mathrm{Z}}_{{\gamma}^{*}}(\hat{\rho})]=1$.  However, this equality together with the inequality in (\ref{2.4}) imply that
\begin{equation}
\parallel \hat{\mathrm{E}}^{\mathrm{X}}_{{\alpha}^{*}}+ \hat{\mathrm{E}}^{\mathrm{Y}}_{{\beta}^{*}}+ \ldots + \hat{\mathrm{E}}^{\mathrm{Z}}_{{\gamma}^{*}} \parallel=n  \label{2.5}
\end{equation}
Since the measurement elements appearing in Eq.~(\ref{2.5}) are positive operators with norms not exceeding unity, we must conclude that they are all in fact of unit norm.  Equation (\ref{2.5}) further implies that for any $\epsilon >0$, there must exist a state ${\hat{\rho}}^{\ast}$ such that $\mid1-\textrm{tr} ( \hat{\mathrm{E}} {\hat{\rho}}^{\ast})\mid < \epsilon$, where $ \hat{\mathrm{E}}$ stands for every measurement element in Eq.~(\ref{2.5}).  But this implies the existence of an approximate common eigenstate contrary to our assumption, thus completing the proof of Theorem 1B.

We are now in a position to summarize the physical content of the uncertainty principle in the framework of majorization theory.

\textbf{The uncertainty principle}.  The joint results of a set of generalized measurements of a given state are no less
uncertain than a probability vector that depends on the measurement set but not the state, and is itself uncertain unless the
measurement elements have a common eigenstate.

In the above statement, we have dropped the qualification ``approximate'' in referring to eigenfunctions since this is the common practice in the physics literature as well as the fact that it makes little difference for the physical content of the uncertainty principle.
\subsection{Quasi-Entropic formulations of uncertainty}
Theorems 1A and 1B immediately imply a parallel formulation of the uncertainty principle for generalized measurements based on quasi-entropic measures.  We recall that a quasi-entropic measure of uncertainty is any symmetric, concave function of the components of ${\mathscr{P}}^{\mathrm{X} \oplus \mathrm{Y} \oplus \ldots \oplus \mathrm{Z}}(\hat{\rho})$ resulting from generalized measurements $\{ \mathrm{X}, \mathrm{Y}, \ldots, \mathrm{Z} \}$ on the state $(\hat{\rho})$ \cite{Schur}.   Specifically, for every quasi-entropic function $F$, we define a scalar uncertainty measure according to
\begin{equation}
\mathcal{U}^{\mathrm{X} \oplus \mathrm{Y} \oplus \ldots \oplus \mathrm{Z}}(F,\hat{\rho})= F[{\mathscr{P}}^{\mathrm{X} \oplus \mathrm{Y} \oplus \ldots \oplus \mathrm{Z}}(\hat{\rho})]-F(\mathcal{I}),  \label{2.6}
\end{equation}
where we have normalized the measure such that it vanishes when measurement results have zero uncertainty and is positive otherwise.

A special class of quasi-entropic measures is obtained if we choose
\begin{equation}
F[{\mathscr{P}}^{\mathrm{X} \oplus \mathrm{Y} \oplus \ldots \oplus \mathrm{Z}}(\hat{\rho})]={\sum}_{\alpha, \beta, \ldots, \gamma} f[\mathscr{P}^{\mathrm{X}}_{{\alpha}}(\hat{\rho})\mathscr{P}^{\mathrm{Y}}_{{\beta}}(\hat{\rho}) \ldots \mathscr{P}^{\mathrm{Z}}_{{\gamma}}(\hat{\rho})] , \label{2.7}
\end{equation}
where $f$ is a concave function of a single variable.  Note that $F$ as constructed in Eq.~(\ref{2.7}) is manifestly symmetric and, as a sum of concave functions, it is also concave.

The standard entropic measure of uncertainty \cite{DEU} corresponds to the choice $f(x)=H(x)=-x \ln(x)$ in Eq.~(\ref{2.7}), in which case the sum on the right-hand side simplifies to the sum of Shannon entropies for each measurement.  This example serves to demonstrate that the class of measures introduced in Eq.~(\ref{2.6}) is a vast generalization of the standard entropic measure of uncertainty, not only in the functional form of the uncertainty measure but also in the fact that it allows for any number of generalized measurements.   Furthermore, as stated in the following formulation of the uncertainty principle based on quasi-entropic functions, the corresponding lower bounds to uncertainty are given by the associated majorization bound.

\textbf{Corollary 1}.   Let $\{ \mathscr{P}^{\mathrm{X}}(\hat{\rho}), \mathscr{P}^{\mathrm{Y}}(\hat{\rho}), \ldots, \mathscr{P}^{\mathrm{Z}}(\hat{\rho}) \}$ be the set of probability vectors resulting from a set of generalized measurements $\{ \mathrm{X}, \mathrm{Y}, \ldots, \mathrm{Z} \}$  on the state $\hat{\rho}$. Then, for any uncertainty measure $\mathcal{U}^{\mathrm{X} \oplus \mathrm{Y} \oplus \ldots \oplus \mathrm{Z}}(F,\hat{\rho})$ as defined in Eq.~(\ref{2.6}), and with ${\mathscr{P}}_{sup}^{\mathrm{X} \oplus \mathrm{Y} \oplus \ldots \oplus \mathrm{Z}}$ as defined in Eq.~(\ref{2.1}), we have
\begin{equation}
\mathcal{U}^{\mathrm{X} \oplus \mathrm{Y} \oplus \ldots \oplus \mathrm{Z}}(F,\hat{\rho}) \geq \mathcal{U}^{\mathrm{X} \oplus \mathrm{Y} \oplus \ldots \oplus \mathrm{Z}}_{min}=F[{\mathscr{P}}_{sup}^{\mathrm{X} \oplus \mathrm{Y} \oplus \ldots \oplus \mathrm{Z}}] > 0, \label{2.8}
\end{equation}
unless the measurement elements $\{ \hat{\mathrm{E}}^{\mathrm{X}}, \hat{\mathrm{E}}^{\mathrm{Y}}, \ldots, \hat{\mathrm{E}}^{\mathrm{Z}} \}$ have an approximate common eigenstate, or a common eigenstate if the elements are compact.   In either of these cases, $\mathcal{U}^{\mathrm{X} \oplus \mathrm{Y} \oplus \ldots \oplus \mathrm{Z}}(F,\hat{\rho})$ vanishes.

This general result is an immediate consequence of Theorems 1A and 1B and the quasi-entropic nature of the underlying measures.  It is important to understand that the uncertainty bound given in Eq.~(\ref{2.8}) is valid, and in fact optimal, for the entire class of quasi-entropic measures.  The optimality property of $F[{\mathscr{P}}_{sup}^{\mathrm{X} \oplus \mathrm{Y} \oplus \ldots \oplus \mathrm{Z}}]$ is a consequence of the optimality of ${\mathscr{P}}_{sup}^{\mathrm{X} \oplus \mathrm{Y} \oplus \ldots \oplus \mathrm{Z}}$, which is in fact the defining characteristic of the latter.  As such, $F[{\mathscr{P}}_{sup}^{\mathrm{X} \oplus \mathrm{Y} \oplus \ldots \oplus \mathrm{Z}}]$ cannot be expected to be optimal for individual members of the quasi-entropic class such as the Shannon or Tsallis measures.

As an example, we will calculate the Shannon-entropic uncertainty bound given by Eq.~(\ref{2.8}) for maximal projective measurements of the three components of a spin-$\frac{1}{2}$ system.  The majorization bound for this measurement, $\mathscr{P}_{sup}^{{\sigma}_{x}\oplus {\sigma}_{y}\oplus {\sigma}_{z} }$, is given in Eq.~(\ref{4.4}) of \S IVB.  Therefore, the desired entropic bound is given by $H(\mathscr{P}_{sup}^{{\sigma}_{x}\oplus {\sigma}_{y}\oplus {\sigma}_{z} })$, so that we can write
\begin{equation}
\mathcal{U}^{{\sigma}_{x}\oplus {\sigma}_{y}\oplus {\sigma}_{z} }(H,\hat{\rho}) \geq H(\mathscr{P}_{sup}^{{\sigma}_{x}\oplus {\sigma}_{y}\oplus {\sigma}_{z} }) =1.23.  \label{2.9}
\end{equation}
The fact that a single majorization bound ${\mathscr{P}}_{sup}^{\mathrm{X} \oplus \mathrm{Y} \oplus \ldots \oplus \mathrm{Z}}$ generates an uncertainty bound for the entire quasi-entropic class of measures is of course a consequence of its power as a comparator of disorder, as discussed in \S I.

In the remainder of this paper we will explore certain consequences of the majorization formulation of the uncertainty principle.  Our objective will be to illustrate the power and reach of the majorization formulation, primarily in applications that are familiar from the traditional variance and entropic formulations of the uncertainty principle.
\section{Mutually unbiased observables}
Our first application is to projective measurements of mutually unbiased observables in two-dimensional Hilbert spaces.
\subsection{Two spin-$\mathbf{\frac{1}{2}}$ components}
The simplest example of Eq.~(\ref{2}) is a maximal projective measurement of a pair of
mutually unbiased observables in a two-dimensional Hilbert space, e.g., a measurement of ${\hat{\sigma}}_{x}$ and
${\hat{\sigma}}_{y}$ on a spin-1/2 system. The state of a spin-1/2 system can in general be represented as $\hat{\rho}=
(1+\mathbb{\hat{\sigma}}\cdot \mathbf{p})/2$, corresponding to a polarization vector $\mathbf{p}$.  The measurement elements, on the other hand, are $\hat{\mathrm{E}}_{1,2}^{{\sigma}_{x}}
=(1 \pm {\hat{\sigma}}_{x})/2$ and $\hat{\mathrm{E}}_{1,2}^{{\sigma}_{y}}=(1\pm {\hat{\sigma}}_{y})/2$.  A calculation using these quantities gives $\mathscr{P}^{{{\sigma}}_{x}}(\hat{\rho})= [(1+{p}_{x})/2, (1-{p}_{x})/2]$ and $\mathscr{P}^{{{\sigma}}_{y}}(\hat{\rho})= [(1+{p}_{y})/2, (1-{p}_{y})/2]$, whereby we find
\begin{align}
\mathscr{P}^{{{\sigma}}_{x}\oplus
{{\sigma}}_{y}}&(\hat{\rho})=[(1+{p}_{x})(1+{p}_{y})/4, (1+{p}_{x})(1-{p}_{y})/4, \nonumber \\
&(1-{p}_{x})(1+{p}_{y})/4,(1-{p}_{x})(1-{p}_{y})/4]. \label{4.1}
\end{align}
The next step is to find the supremum of $\mathscr{P}^{{{\sigma}}_{x}\oplus {{\sigma}}_{y}}(\hat{\rho})$ as $\hat{\rho}$, or equivalently
$\mathbf{p}$, is varied.  Following the construction of Eq.~(\ref{1.5}) \textit{et} \textit{seq}., we first determine
${\mu}^{sup}_{1}$ by finding the maximum value of a single component of $\mathscr{P}^{{{\sigma}}_{x}\oplus {{\sigma}}_{y}}(\hat{\rho})$ in Eq.~(\ref{4.1}),
then ${\mu}^{sup}_{2}$ by finding the maximum value of the sum of a pair of components of $\mathscr{P}^{{{\sigma}}_{x}\oplus
{{\sigma}}_{y}}(\hat{\rho})$, and so on.   Implementing this process, we find
\begin{equation}
{\mu}^{sup}=[0, {(1+1/\sqrt{2})}^{2}/4,1,1,1], \label{4.2}
\end{equation}
where ${\mu}^{sup}_{1}$ obtains for ${p}_z=0$ and $|\mid {p}_x|\mid=\mid{p}_y\mid=1/\sqrt{2}$, and the next three components for $\textbf{p}$ equal
to a unit vector along either the x- or y-axis. Using Eq.~(\ref{4.2}) and the counterpart of Eq.~(\ref{1.5}) for the supremum, we arrive at
\begin{equation}
\mathscr{P}_{sup}^{{{\sigma}}_{x}\oplus
{{\sigma}}_{y} }=[(1.5+\sqrt{2})/4, (2.5-\sqrt{2})/4,0,0].  \label{4.3}
\end{equation}
Thus a measurement of ${\hat{\sigma}}_{x}$ and ${\hat{\sigma}}_{y}$ on any
spin-1/2 system will yield results no more certain than this supremum.  Notice that while the supremum in Eq.~(\ref{4.3}) is less uncertain than is possible for any state of the system, it is not itself a possible probability vector in any actual measurement \cite{note1}.  This is therefore an instance of a measurement where there is no ``minimum uncertainty'' state in the majorization sense, as discussed in \S IIIA.

We note in passing here that the infimum for the above measurement, $\mathscr{P}_{inf}^{{{\sigma}}_{x}\oplus
{{\sigma}}_{y} }$, is trivially realized on an unpolarized state (i.e., for $\mathbf{p}=0$) with all four components equal $1/4$.  As such, it represents the state of maximum uncertainty for the measurement. If $\hat{\rho}$ is restricted to pure states, on the other hand, the supremum is still given by Eq.~(\ref{4.3}) but the infimum is found to be $(1/2,1/2,0,0)$, which is realized when $\hat{\rho}$ is an eigenstate of one or the other of the two observables.
\subsection{Three spin-$\mathbf{\frac{1}{2}}$ components}
The extension of the above analysis to the case of three mutually unbiased observables, e.g., all three components of
$\boldsymbol{\hat{\sigma}}$ in the foregoing example, is analogous but requires the full machinery of the calculation of the supremum.  Here the measurement set is ${{\sigma}_{x}\oplus {\sigma}_{y}\oplus {\sigma}_{z}}$, with the six measurement elements given by $\hat{\mathrm{E}}_{1,2}^{{\sigma}_{x}}
=(1 \pm {\hat{\sigma}}_{x})/2$, $\hat{\mathrm{E}}_{1,2}^{{\sigma}_{y}}=(1\pm {\hat{\sigma}}_{y})/2$, and $\hat{\mathrm{E}}_{1,2}^{{\sigma}_{z}}=(1\pm {\hat{\sigma}}_{z})/2$.  The state is parametrized as above, so that $\hat{\rho}= (1+\mathbb{\hat{\sigma}}\cdot \mathbf{p})/2$.  A straightforward calculation of the eight components of $\mathscr{P}^{{\sigma}_{x}\oplus {\sigma}_{y}\oplus {\sigma}_{z}}(\hat{\rho})$ now gives $(1\pm{p}_{x})(1\pm{p}_{y})(1\pm{p}_{z})/8$, which extends the result given in Eq.~(\ref{4.1}) to three observables.

The procedure for finding $\mathscr{P}_{sup}^{{\sigma}_{x}\oplus {\sigma}_{y}\oplus {\sigma}_{z} }$ is the same as above, i.e., maximizing a single element of $\mathscr{P}^{{\sigma}_{x}\oplus {\sigma}_{y}\oplus {\sigma}_{z}}(\hat{\rho})$, then the sum of a pair of components, and so on, followed by the ``flattening'' process described in \S IIB to obtain a descending sequence.  While doable analytically, this calculation is more conveniently done numerically \cite{code1}.   Using either method, one finds for the first two components of ${\mu}^{sup}$ the values ${\mu}^{sup}_{1}={(1+1/\sqrt{3})}^{3}/8$ and ${\mu}^{sup}_{2}={(1+1/\sqrt{2})}^{2}/4$.  The next two components emerge in an ascending order and must therefore be flattened, i.e., replaced by their mean.   It turns out that these four components add up to unity, thus implying that the next four components vanish.  The desired supremum is then found from ${\mu}^{sup}$ and is given by
\begin{equation}
\mathscr{P}_{sup}^{{\sigma}_{x}\oplus {\sigma}_{y}\oplus {\sigma}_{z} }=(0.491,0.238,0.136,0.136,0,0,0,0). \label{4.4}
\end{equation}
Thus any measurement of the three components of a spin-$\frac{1}{2}$ system will yield results more uncertain than the supremum given in Eq.~(\ref{4.4}), while the latter itself cannot be reached in any physically realizable measurement.

As in the case of two spin components, $\mathscr{P}_{inf}^{{{\sigma}}_{x}\oplus
{{\sigma}}_{y} \oplus {{\sigma}}_{z} }$ is trivially realized on an
unpolarized state with all eight components equal.  With $\hat{\rho}$ restricted to pure states, the supremum is unchanged while the
infimum is found to be $(0.250,0.250,0.250,0.104,0.062,0.040,0.034,0.011)$.

Needless to say, the results given in Eqs.~(\ref{4.3}) and (\ref{4.4}) conform to the requirements of Theorem 1A.
\section{Canonically conjugate observables}
Here we consider position and momentum, the archetypal example of the uncertainty principle for canonically conjugate observables.  Our objective is to calculate the leading component of
${\mathscr{P}}_{sup}^{{x} \oplus {p}}$ for a projective measurement of position and momentum in one dimension since the knowledge of this component is sufficient to determine whether ${\mathscr{P}}_{sup}^{{x} \oplus {p}} \prec \prec \mathcal{I}$ as required by Theorem 1B.  As expected, we will find that this condition is fulfilled in this case as well.

A projective measurement of position in one dimension, $\hat{x}$, entails a set of measurement bins corresponding to intervals
of the x-axis (where the detectors are positioned) \cite{HP1}.  Let $[{x}_{1,\alpha}, {x}_{2,\alpha}]$ be the $\alpha$th bin, with
${\hat{\Pi}}^{{x}}_{\alpha}$ the corresponding Hilbert space projection operator, and similarly $ {\hat{\Pi}}^{{p}}_{\beta}$ and
$[{p}_{1,\beta}, {p}_{2,\beta}]$ for momentum ($\hat{p}$) measurement \cite{note3}.  Note that these projection operators are the measurement elements for this measurement set, i.e., $\hat{\mathrm{E}}^{{x}}_{\alpha}= {\hat{\Pi}}^{{x}}_{\alpha}$ and $\hat{\mathrm{E}}^{{p}}_{\beta}= {\hat{\Pi}}^{{p}}_{\beta}$.   The explicit representation of these projection operators in coordinate space are as follows:
\begin{align}
\langle x \mid {\hat{\Pi}}^{{x}}_{\alpha} \mid x' \rangle &= \delta(x-x') \Theta({x}_{1,\alpha}-x)\Theta(x-{x}_{2,\alpha}), \nonumber \\
\langle x \mid {\hat{\Pi}}^{{p}}_{\beta} \mid x' \rangle &=\frac{1}{2\pi} \exp[i {\bar{p}}_{\beta}(x-x')] \frac{\sin[{\Delta p}_{\beta}(x-x')/2] }{(x-x')/2},  \label{5.01}
\end{align}
where ${\bar{p}}_{\beta}=({p}_{2,\beta}+{p}_{1,\beta})/2$ and ${\Delta p}_{\beta}={p}_{2,\beta}-{p}_{1,\beta}$.  We have also set $\hbar =1$ in Eq.~(\ref{5.01}) to simplify the writing.

Our objective is to find the maximum value of ${\mathscr{P}}_{\alpha \beta}^{{x} \oplus {p}}$, which by definition equals ${\mathscr{P}}^{{x}}_{\alpha}(\hat{\rho}) {\mathscr{P}}^{{p}}_{\beta}(\hat{\rho}) = \textrm{tr} ({\hat{\Pi}}^{{x}}_{\alpha} \hat{\rho}) \textrm{tr} ({\hat{\Pi}}^{{p}}_{\beta}
\hat{\rho})$, as $\hat{\rho}$ is varied.  Since $\hat{\rho}$ is a convex mixture of pure states, the desired maximum will be realized on pure
states for which $\hat{\rho}\rightarrow |\psi \rangle \langle \psi |$.  We are thus looking to maximize $\langle \psi
|{\hat{\Pi}}^{{x}}_{\alpha} |\psi \rangle \langle \psi |{\hat{\Pi}}^{{p}}_{\beta} |\psi \rangle / {\langle \psi |\psi \rangle}^2$ by varying $|\psi \rangle$, or
equivalently $\langle \psi |$.

A variation with respect to $\langle \psi |$ gives the eigenvalue equation
\begin{equation}
{\hat{\cal{L}}}_{+} |{\psi}^{\star} \rangle= |{\psi}^{\star} \rangle, \label{5.1}
\end{equation}
where
\begin{equation}
{\hat{\cal{L}}}_{\pm}= ({\hat{\Pi}}^{{x}}_{\alpha}/{\mathscr{P}}^{{x \star} }_{\alpha} \pm {\hat{\Pi}}^{{p}}_{\beta}/{\mathscr{P}}^{{p \star } }_{\beta})/2, \label{5.2}
\end{equation}
and where we have used a star to signify an optimized quantity so that ${\mathscr{P}}^{{x \star} }_{\alpha}= \langle {\psi}^{\star}|{\hat{\Pi}}^{{x}}_{\alpha} |{\psi}^{\star}\rangle /{\langle {\psi}^{\star} |{\psi}^{\star} \rangle}$ and ${\mathscr{P}}^{{p \star } }_{\beta}= \langle {\psi}^{\star}|{\hat{\Pi}}^{{p}}_{\beta} |{\psi}^{\star}\rangle /{\langle {\psi}^{\star} |{\psi}^{\star} \rangle}$.   The two operators ${\hat{\cal{L}}}_{\pm}$ defined above are clearly bounded and self-adjoint, with ${\hat{\cal{L}}}_{+}$ positive as well.  Furthermore, ${\hat{\cal{L}}}_{-}$ has a vanishing expectation value in the optimized state $|{\psi}^{\star} \rangle$ by definition.

Using Eq.~(\ref{5.1}) and the fact that $\langle {\psi}^{ \star} |{\hat{\cal{L}}}_{-} |{\psi}^{ \star} \rangle =0$, we find that $\langle {\psi}^{ \star} |{\hat{\cal{L}}}_{+}{\hat{\cal{L}}}_{-}+{\hat{\cal{L}}}_{-}{\hat{\cal{L}}}_{+} |{\psi}^{ \star} \rangle =0$.  If we then apply the definitions of ${\mathscr{P}}^{{x \star}}_{\alpha}$ and ${\mathscr{P}}^{{p \star} }_{\beta}$ to this equation, we arrive at the equality ${\mathscr{P}}^{{x \star}}_{\alpha}={\mathscr{P}}^{{p \star} }_{\beta}$.  This equality is a consequence of the symmetry with respect to the $\hat{x}\leftrightarrows \hat{p}$ exchange in the above optimization problem.

At this point we appeal to Eq.~(\ref{2.4}) of \S IIIB, which implies that ${\mathscr{P}}^{{x}}_{\alpha}(\hat{\rho}) {\mathscr{P}}^{{p}}_{\beta}(\hat{\rho})\leq \frac{1}{4}{\parallel {\hat{\Pi}}^{{x}}_{\alpha} + {\hat{\Pi}}^{{p}}_{\beta}\parallel}^{2}$. This inequality, in view of ${\mathscr{P}}^{{x \star}}_{\alpha}={\mathscr{P}}^{{p \star} }_{\beta}$, in turn implies that ${\mathscr{P}}^{{x \star}}_{\alpha}={\mathscr{P}}^{{p \star} }_{\beta}\leq \frac{1}{2}{\parallel {\hat{\Pi}}^{{x}}_{\alpha} + {\hat{\Pi}}^{{p}}_{\beta}\parallel}$.   Comparing this to Eq.~(\ref{5.1}), we conclude that
\begin{equation}
{\mathscr{P}}^{{x \star}}_{\alpha}={\mathscr{P}}^{{p \star} }_{\beta} = \frac{1}{2}{\parallel {\hat{\Pi}}^{{x}}_{\alpha} + {\hat{\Pi}}^{{p}}_{\beta}\parallel}.  \label{5.2}
\end{equation}
Thus the eigenvalue of unity in Eq.~(\ref{5.1}) is in fact the maximum for the operator ${\hat{\cal{L}}}_{+} $.  Our task then is to find ${\parallel {\hat{\Pi}}^{{x}}_{\alpha} + {\hat{\Pi}}^{{p}}_{\beta}\parallel}$.

To that end, we left-multiply Eq.~(\ref{5.1}) by ${\hat{\cal{L}}}_{+} $, use Eq.~(\ref{5.1}) again together with the idempotent property of the projection operators to eliminate all operators except ${\hat{\Pi}}^{{p}}_{\beta}{\hat{\Pi}}^{{x}}_{\alpha}$, and rewrite the resulting equation in terms of $|\phi ^{ \star}\rangle = {\hat{\Pi}}^{{x}}_{\alpha}|{\psi}^{ \star} \rangle$.  The result is the transformed equation
\begin{equation}
{\hat{\Pi}}^{{x}}_{\alpha}{\hat{\Pi}}^{{p}}_{\beta}{\hat{\Pi}}^{{x}}_{\alpha}\mid \phi ^{ \star}\rangle ={(2{\mathscr{P}}^{{x \star}
}_{\alpha}-1)}^{2} |\phi ^{ \star}\rangle.   \label{5.25}
\end{equation}
Thus ${(2{\mathscr{P}}^{{x \star} }_{\alpha}-1)}^{2}$ equals the largest eigenvalue of the positive operator ${\hat{\Pi}}^{{x}}_{\alpha}{\hat{\Pi}}^{{p}}_{\beta}{\hat{\Pi}}^{{x}}_{\alpha}$, which we denote by ${\mu}_{max}^{2}$. Consequently, the desired maximum, ${\mathscr{P}}_{\alpha \beta}^{{x} \oplus {p} \star}={({\mathscr{P}}^{{x \star} }_{\alpha})}^{2}$, equals $\frac{1}{4}{(1+{\mu}_{max})}^2$.  At this point we observe that $\parallel {\hat{\Pi}}^{{x}}_{\alpha} {\hat{\Pi}}^{{p}}_{\beta}{\hat{\Pi}}^{{x}}_{\alpha}\parallel \leq 1$ since it involves a product of projection operators.  This in turn implies that ${\mu}_{max} \leq 1$ and consequently ${\mathscr{P}}_{\alpha \beta}^{{x} \oplus {p} \star} \leq 1$ as well.  Furthermore, ${\mathscr{P}}_{\alpha \beta}^{{x} \oplus {p} \star} = 1$ is excluded since it would imply ${\parallel {\hat{\Pi}}^{{x}}_{\alpha} + {\hat{\Pi}}^{{p}}_{\beta}\parallel}=2$ by Eq.~(\ref{5.2}) and an approximate common eigenstate for $\parallel {\hat{\Pi}}^{{x}}_{\alpha} \parallel$ and $\parallel {\hat{\Pi}}^{{p}}_{\beta}\parallel$ which is not possible for finite $x$ and $p$ bins.

To simplify the notation, we can without a loss of generality translate the $x$ and $p$ axes such the two bins $\alpha$ and $\beta$ for which the optimal values above are reached are symmetrically centered at $x=0$ and $p=0 $ with  ${x}_{2,\alpha}=-{x}_{1,\alpha}=\Delta x/2$ and $ {p}_{2,\beta}=-{p}_{1,\beta}=\Delta p /2$.  Using the representations of the projection operators given in Eq.~(\ref{5.01}), we can write Eq.~(\ref{5.25}) as the following integral equation:
\begin{equation}
\frac{1}{2\pi} {\int}_{-\Delta x /2 }^{+\Delta x /2 }  dx' \frac{\sin[{\Delta p}_{\beta}(x-x')/2] }{(x-x')/2} \phi ^{ \star}(x')={\mu}_{max}^{2}\phi ^{ \star}(x), \label{5.3}
\end{equation}
where $\phi ^{ \star}(x)= \langle x \mid \phi ^{ \star}\rangle$ is the wavefunction corresponding to the state $|\phi ^{ \star}\rangle = {\hat{\Pi}}^{{x}}_{\alpha}|{\psi}^{ \star} \rangle$, with $x$ and $x'$ restricted to the interval $[-\Delta x/2, +\Delta x/2]$.  It is convenient to rescale Eq.~(\ref{5.3}) by measuring $x$ and $x'$ in units of $\Delta x$.  The resulting equation can then be written as
\begin{equation}
\frac{1}{\pi}{\int}_{-1/2 }^{+1/2 }  d\xi' \frac{\sin[s\pi(\xi - \xi')] }{(\xi - \xi')} f(\xi')={{\mu}^{2}}f(\xi), \label{5.4}
\end{equation}
where $s=(\Delta x)(\Delta p)/2\pi \hbar$ and $\xi$ and $\xi'$ are restricted to the interval $[-1/2, +1/2]$.  Note that $\hbar$ has been restored to the expression for $s$ here.

The largest eigenvalue of the integral equation (\ref{5.4}) is thus equal to ${{\mu}_{max}^{2}}$.  Its kernel, on the other hand, is a positive operator bounded by unity according to Eq.~(\ref{5.2}) \textit{et seq}.  In addition, the square of this operator has a finite trace, which implies that the kernel belongs to the Hilbert-Schmidt class of operators and is therefore compact as well \cite{ReSi}.  This confirms that the spectrum of Eq.~(\ref{5.4}) is discrete and of finite multiplicity (except possibly for zero), confined to the interval from zero to one, and can only cluster around zero.  Furthermore, the sum of the eigenvalues of the operator, which may be found by calculating its trace, equals $s$.

The spectrum of Eq.~(\ref{5.4}) can be intuitively captured by considering $(\Delta x \Delta p)$ as the ``volume of phase space'' and of $s=(\Delta x)(\Delta p)/2\pi \hbar$ as the ``number of states'' as well as the number of (non-zero) eigenvalues.  For a large phase space volume, there are $s$ eigenstates with non-zero eigenvalues nearly equal to unity so that their sum should be of the order of $s$, which agrees with the trace of the kernel found above.  Indeed a perturbative treatment of Eq.~(\ref{5.4}) for large values of $s$ confirms this interpretation \cite{pert}.  Indeed for $s \rightarrow \infty$, we find, using the stationary phase approximation, that the kernel of Eq.~(\ref{5.4}) effectively approaches $\delta(\xi - \xi')$ and ${{\mu}^{2}}={{\mu}^{2}}_{max} \rightarrow 1$.  This is the limit of low-precision measurements with position or momentum bins approaching the entire span of the $x$ or $p$ space and ${\mathscr{P}}_{\alpha \beta}^{{x} \oplus {p} \star}\rightarrow 1$.  Therefor the measurement results approach (but do not reach) the limit of zero uncertainty for large $s$, albeit for measurements that yield correspondingly little information.  The physically interesting limit is of course the opposite extreme of small bins and high resolution where the effects of the uncertainty principle are most strongly manifested.  We will next consider that limit.

For small bins and high resolution measurements, we expect no more than one small non-zero eigenvalue for Eq.~(\ref{5.4}).  To verify this expectation, we observe that as $s \rightarrow 0$, ${\sin[s\pi(\xi - \xi')] }/{(\xi - \xi')} \rightarrow s \pi$, $u(\xi)\rightarrow  \theta (1-4{\xi}^2)$ (arbitrary normalization), and ${{\mu}^{2}} ={{\mu}^{2}}_{max}\rightarrow s $, where $\theta $ is the usual step function.  Thus the largest eigenvalue approaches the sum of all eigenvalues, confirming that there is just one non-zero eigenvalue in this limit.  Using its value, we find ${({\mathscr{P}}^{{x \star} }_{\alpha})} \rightarrow  \frac{1}{2}{(1+\sqrt{s})} $ as $s \rightarrow 0$, and therefor,
\begin{equation}
{\mathscr{P}}_{\alpha \beta}^{{x} \oplus {p} \star}  \stackrel{s \rightarrow 0}{\longrightarrow } \frac{1}{4}{(1+2\sqrt{s})}. \label{5.5}
\end{equation}

The corresponding wavefunction, ${\psi}^{\star}(x)= \langle x \mid{\psi}^{\star}\rangle$, can be constructed by reference to the two projections ${\phi}^{\star}(x)=\langle x\mid {\hat{\Pi}}^{{x}}_{\alpha}|{\psi}^{ \star} \rangle$ and ${\chi}^{\star}(x)=\langle x\mid {\hat{\Pi}}^{{p}}_{\beta}|{\psi}^{ \star} \rangle$.   The first of these was defined above and is directly related to $u(\xi)\rightarrow  \theta (1-4{\xi}^2)$ as found in the foregoing paragraph.  The second, ${\chi}^{\star}(x)$, can be constructed from the first using the $\hat{x}\leftrightarrows \hat{p}$ symmetry mentioned above.   Omitting the details of these steps, we can state the result as follows:
\begin{equation}
{\psi}^{ \star}(x) \stackrel{s \rightarrow 0}{\longrightarrow }\frac{1}{\sqrt{2 \Delta x}} \theta ({\Delta x}^{2}-4x^2) +\sqrt{\frac{\Delta p}{4\pi \hbar} } \frac{\sin(x \Delta p /2 \hbar)}{(x \Delta p /2 \hbar)}. \label{5.6}
\end{equation}
In view of the $\hat{x}\leftrightarrows \hat{p}$ symmetry mentioned above, it is instructive to consider the momentum-space representation of this equation, ${\tilde{\psi}}^{ \star}(p)=\langle p \mid{\psi}^{\star}\rangle$, which can be found by Fourier transformation:
\begin{equation}
{\tilde{\psi}}^{ \star}(p) \stackrel{s \rightarrow 0}{\longrightarrow }\frac{1}{\sqrt{2 \Delta p}} \theta ({\Delta p}^{2}-4p^2) +\sqrt{\frac{\Delta x}{4\pi \hbar} } \frac{\sin(p \Delta x /2 \hbar)}{(p \Delta x /2 \hbar)}. \label{5.7}
\end{equation}
The expected symmetry is clearly in evidence between Eqs.~(\ref{5.6}) and (\ref{5.7}).  It is important to remember here that $\Delta x$ and $\Delta p$ are bin sizes and in effect represent the resolution of the measuring devices.  They should not be confused with variances.

The above results, i.e., ${({\mathscr{P}}^{{x \star} }_{\alpha})} = {({\mathscr{P}}^{{p \star} }_{\beta})} \rightarrow  \frac{1}{2}{(1+\sqrt{s})} $ as $s \rightarrow 0$, show that \textit{the state with the sharpest simultaneous values of position and momentum will turn up the two values only $25\%$ of the time}.  Note also that as a characterization of the limitations on what is knowable in a measurement of position and momentum, this is a more poignant statement of the uncertainty principle than the Heisenberg inequality or the optimal entropic bound \cite{BIA}.  This is so even though we have only used the leading component of ${\mathscr{P}}^{{x} \oplus {p}\star }$ for its derivation.  As pointed out earlier, $\mid{\psi}^{\star}\rangle$ is not a ``minimum uncertainty'' state in the usual sense as it only maximizes the leading component of the probability vector. The calculation of the remaining terms of ${\mathscr{P}}^{{x} \oplus {p} \star}$ in the high-precision limit, which will impose limitations on the readouts of more than one position or momentum bin at a time, is an interesting unsolved problem.
\section{Least uncertain measurement of a state}
Given a state $\hat{\rho}$, different measurements performed on it will give
rise to probability vectors of varying uncertainty.  Is there a measurement ${\mathrm{X}}^{\star}$ which results in a minimally
uncertain probability vector such that $\mathscr{P}^{\mathrm{X}}(\hat{\rho}) \prec
\mathscr{P}^{{\mathrm{X}}^{\star}}(\hat{\rho})$ for any measurement ${\mathrm{X}}$?   This question is of course only meaningful if restricted to measurements of comparable precision, and interesting only if applied to high-precision measurements such as those that are rank-1, since low-precision measurements can yield probability vectors of arbitrarily low uncertainty. We will therefore look for ${\mathrm{X}}^{\star}$ among rank-1 measurements.  To that end, we will first develop a sharpened version of Eq.~(29) of Ref.~\cite{NIE}.

\textbf{Lemma}.  Suppose a rank-1 measurement ${\mathrm{X}}$ is performed on a state $\hat{\rho}$ resulting in states $\{
{\hat{\rho} '}_{\alpha} \}$ with probabilities $\{ \mathscr{P}_{\alpha}^{{\mathrm{X}}} \}$. We then have (i) ${\oplus}_{\alpha}\, \mathscr{P}_{\alpha}^{{\mathrm{X}}}\, {\lambda}({\hat{\rho} '}_{\alpha})\prec {\lambda}({\hat{\rho}})$, where
\begin{equation}
{\sup}_{\,\mathrm{X}}[ {\oplus}_{\alpha}\, \mathscr{P}_{\alpha}^{{\mathrm{X}}}\, {\lambda}({\hat{\rho} '}_{\alpha})]={\lambda}^{\downarrow}({\hat{\rho}}), \label{7}
\end{equation}
and (ii) there exits a rank-1 projective measurement ${{{\mathrm{X}}^{\star}}}$ that realizes the above supremum.

Above, ${\lambda}^{\downarrow}(\cdot)$ is the spectrum of a density matrix in a descending order, and ``$\oplus$'' denotes a direct sum
of spectra as defined in Ref.~\cite{NIE}.  The majorization condition in part (i) of this lemma was proved in Ref.~\cite{NIE}; cf. Eq.~(29) therein.

For the equality in part (i), as well as part (ii), we only need to demonstrate the existence of a rank-1 measurement that equals the right-hand side of Eq.~(\ref{7}), thus realizing the supremum.   One can readily verify that the projective measurement
defined by $ {\hat{\Pi}}_{\alpha}^{\star}=\mid \alpha \rangle \langle \alpha \mid$, where $\{\mid \alpha \rangle \langle \alpha \mid\}$ are
the eigenstates of $\hat{\rho}$ arranged according to descending eigenvalues, is in fact the desired measurement ${\mathrm{X}}^{\star}$.  This is because
with ${\mathrm{X}}$ restricted to rank-1 measurements, generalized or projective, every ${\hat{\rho}'}_{\alpha}$ is pure so that ${\lambda}^{\downarrow}({\hat{\rho}'}_{\alpha})$ equals
$(1,0, \ldots,0)$ and ${\oplus}_{\alpha}\, \mathscr{P}_{\alpha}^{{\mathrm{X}}}\, \lambda({\hat{\rho}'}_{\alpha})
=\mathscr{P}^{{\mathrm{X}}}(\hat{\rho})$.   For ${{{\mathrm{X}}^{\star}}}$, on the other hand, $\mathscr{P}^{{\mathrm{X}^{\star}}}_{\alpha}(\hat{\rho})=\textrm{tr}(\mid \alpha \rangle \langle \alpha \mid \hat{\rho})={\lambda}^{\downarrow}_{\alpha}(\hat{\rho})$.   Putting the last two equalities together, we arrive at
\begin{equation}
{\oplus}_{\alpha}\, \mathscr{P}_{\alpha}^{{\mathrm{X}}^{\star}} \lambda({\hat{\rho}'}_{\alpha})={\lambda}^{\downarrow}_{\alpha}(\hat{\rho}), \label{7.1}
\end{equation}
showing that measurement ${{{\mathrm{X}}^{\star}}}$ realizes the supremum in Eq.~(\ref{7}).  This proves the lemma, which allows us to state the following theorem.

\textbf{Theorem 2}. The probability vector resulting from a rank-1 generalized measurement of a state is majorized by the spectrum of that
state, $\mathscr{P}^{{\mathrm{X}}}(\hat{\rho}) \prec {\lambda}(\hat{\rho})$.  Furthermore, for any quasi-entropic measure $F$, $F[{\lambda}(\hat{\rho})] \leq  F[\mathscr{P}^{{\mathrm{X}}}(\hat{\rho})]$.  In particular, the von Neumann entropy of a state is the infimum of the Shannon entropy of all rank-1 measurements of that state, i.e., $S(\hat{\rho}) \leq H[\mathscr{P}^{{\mathrm{X}}}(\hat{\rho})]$.  Furthermore, there is a projective measurement ${\mathrm{X}}^{\star}$ which satisfies $\mathscr{P}^{{\mathrm{X}^{\star}}}(\hat{\rho})={\lambda}(\hat{\rho})$ and saturates the stated inequalities.

The second conclusion of Theorem 2 follows from the fact that ${\lambda}^{1} \prec {\lambda}^{2}$ implies $F({\lambda}^{1}) \geq
F({\lambda}^{2})$ if $F$ is a quasi-entropic measure (i.e., if it is a symmetric, concave function).  The choice of the Shannon entropy function $H$ for F then yields the next part of Theorem 2. This part was also established in the first paper of Ref.~\cite{HP2} for projective measurements.

Theorem 2 provides a fundamental characterization of the von Neumann entropy as the greatest lower bound of all rank-1 measurement (Shannon) entropies.  The majorization statement of the theorem, on the other hand, provides an operational meaning for the spectrum of a quantum state (which is the set of eigenvalues of its density operator) as the least uncertain probability vector resulting from rank-1 measurements of that state.  Not surprisingly, the corresponding least uncertain measurement ${\mathrm{X}}^{\star}$, which saturates all majorization and quasi-entropic relations above, reproduces the pre-measurement state on average, i.e., in the absence of post-selection.

Note that in the trivial case of a pure state, $\hat{\rho}=\mid \psi \rangle$, we find ${\lambda}(\mid \psi \rangle)=(1,0, \ldots,0)$ and $\mathscr{P}^{{\mathrm{X}^{\star}}}(\mid \psi \rangle)=\mathcal{I}$, corresponding to zero uncertainty.  In this case, the optimal measurement ${\mathrm{X}}^{\star}$ is given by the orthogonal projections $ {\hat{\Pi}}_{1}^{\star}= \mid \psi \rangle \langle \psi \mid$ and $ {\hat{\Pi}}_{i}^{\star}=\mid i \rangle \langle i \mid$, $i=2,3, \ldots$, where the states $\{ \mid i \rangle \}$ together with $\mid \psi \rangle$ form a complete orthonormal set.  In case of a pure state, then, the least uncertain rank-1 measurement reproduces the pre-measurement state with certainty.  This amounts to a full identification of the state of the quantum system, which embodies all available information about the system, and zero uncertainty.  In case of a mixed state, on the other hand, the optimal measurement ${\mathrm{X}}^{\star}$ turns up the uncertainties that result from the impurity of the quantum state.  In either case, ${\mathrm{X}}^{\star}$ embodies all available information about the system, as would be expected of a least uncertain measurement of highest resolution.
\section{concluding remarks}
It worth recalling here that majorization is a more refined comparator of uncertainty than those based on a scalar condition, and that the corresponding formulation of the uncertainty principle is more stringent than the quasi-entropic or variance formulations.  We have explained the reasons that set the majorization order apart from the others, and have demonstrated the consequences in several instances.  The price of this generality is a mathematically more complex scheme, as one should expect in going from a scalar to a vector formulation.   Generally speaking, one may expect the majorization formulation to be most suitable when dealing with overarching information-theoretical aspects of quantum systems, such as in Corollary 1 and \S V and VI.  As such, it is a useful complement to the entropic and variance formulations.

In developing majorization bounds, we resorted to the concept of the supremum of a set of probability vectors.  It is important to distinguish the supremum from a \textit{maximal element} for such a set, which would be defined as an element that is not majorized by any other element in the set.   A set of probability vectors may include many maximal elements but no supremum, i.e., the supremum exists but may not be a member of the set.  This is the reason why there is in general no ``minimum uncertainty'' state within the majorization formulation.

In this paper we have presented a few basic applications of the new formulation, with many others remaining to be worked out.  A challenging case is the calculation of the non-leading components of ${\mathscr{P}}_{\alpha \beta}^{{x} \oplus {p} \star}$ for position and momentum measurements.  Another interesting application currently under development is the use of majorization uncertainty bounds for entanglement detection.\\

\begin{acknowledgments}
I would like to thank Y. Huang for useful comments.  This work was in part supported by a grant from California State University, Sacramento.
\end{acknowledgments}

%\begin{references}
{}


\begin{thebibliography}{}
\bibitem{state}  In this paper we will use a caret to distinguish operators, and assume all density operators such as $\hat{\rho}$ to have unit trace.
\bibitem{rank} The rank of an operator is the dimension of its range, so that ${\hat{\mathfrak{M}}}$ is rank-1 if it has the structure $c \mid\phi \rangle \langle \psi\mid$, where $c$ is a complex number.  Clearly, ${\hat{\mathfrak{M}}}\hat{ \rho} {\hat{\mathfrak{M}}}^{\dagger}$ will then equal a multiple of the pure state $\mid\phi \rangle \langle \phi \mid$ for any $\hat{\rho}$, pure or mixed.
\bibitem{note1} The infimum or supremum of a (finite or infinite) set of vectors always exists but need not be a member of the set \cite{note11}.
\bibitem{note11}  Here we have stated the definition of the infimum for a set of $N$ vectors.  In case the number of vectors is infinite, the function ``min'' must be replaced with ``glb'' (greatest lower bound).  Similarly, ``max'' must be replaced with ``lub'' (least upper bound) in the definition of the supremum.  In such cases we would be dealing with the greatest lower, or the least upper, bound of a bounded set of real numbers which always exists.
\bibitem{eig} Specifically, the spectrum of a non-compact Hermitian operator such as a projection operator onto an interval of position or momentum space (treated in \S V) may include a continuous component which would preclude the existence of mathematically well-defined eigenstates or eigenvalues; see Ref. \cite{ReSi}.  A familiar example of this mathematical difficulty is the position operator which does not have proper eigenvalues or eigenstates, even if the system is confined to a finite region of space.
\bibitem{note12} The largest eigenvalue of a compact oerator equals its norm and has finite multiplicity, exactly as in the case of a finit-dimensional hermitian matrix.  Therefore, an eigenstate always exists that has the norm as its eigenvalue.
\bibitem{code1}  Numerical calculations were done by using the code given in \cite{code} and searching over the magnitude and direction of the polarization vector $\mathbf{p}$.
\bibitem{note3}  Note that, our notation notwithstanding, at least one measurement bin must be infinite in size, e.g., $|x|\geq L$, to account for position outcomes outside the apparatus boundaries, where $L$ is of the order of the apparatus size.
\end{thebibliography}
\end{document}